\documentclass[11pt]{article}
\usepackage{epsfig}
\usepackage{dcolumn}
\usepackage{bm}
\usepackage{graphicx}
\usepackage[textsize=tiny]{todonotes}
\usepackage{subcaption}
\usepackage{here}
\usepackage{amssymb,amsmath}
\usepackage{multirow}
\usepackage{cite,color,url}
\usepackage{tabu}
\usepackage{array}
\usepackage[colorlinks=true,urlcolor=blue,anchorcolor=blue
,citecolor=blue,filecolor=blue,linkcolor=blue,menucolor=blue
,linktocpage=true,pdfproducer=medialab,pdfa=true]{hyperref}

\usepackage{colordvi}
\usepackage{epsfig,psfrag,rotating,soul}
\def\beqa{\begin{eqnarray}}
\def\eeqa{\end{eqnarray}}

\evensidemargin \oddsidemargin
\allowdisplaybreaks

\psfrag{lk}{$l_k$}
\psfrag{lm}{$l_m$}
\psfrag{blk}{$\bar {l}_k$}
\psfrag{blm}{$\bar {l}_m$}
\def\thefootnote{\fnsymbol{footnote}}

\let\OLDthebibliography\thebibliography
\renewcommand\thebibliography[1]{
\OLDthebibliography{#1}
\setlength{\parskip}{0pt}
\setlength{\itemsep}{0pt plus 0.3ex}}
\usepackage{makecell}

\newcommand{\res}{R_\sigma}
\newcommand{\ims}{I_\sigma} 
\newcommand{\red}{R_\phi}
\newcommand{\imd}{I_\phi} 
\newcommand{\ret}{R_\Delta}
\newcommand{\imt}{I_\Delta} 

\newcommand{\ott}{123-}
\newcommand{\VHiggs}{V(\sigma,\phi,\Delta)}
\newcommand{\musig}{\mu_\sigma}
\newcommand{\muphi}{\mu_\phi}
\newcommand{\mudel}{\mu_\Delta}
\newcommand{\vs}{v_\sigma}
\newcommand{\vd}{v_\phi}
\newcommand{\vt}{v_\Delta}

\newcommand{\matr}{\mathcal}
\newcommand{\Ueven}{{\sc U}^H}
\newcommand{\Uodd}{{\sc U}^A}
\newcommand{\UChar}{{\sc U}^C}
\newcommand{\cp}{\rm CP}
\newcommand{\beq}{\begin{eqnarray}}
\newcommand{\eeq}{\end{eqnarray}}

\newcommand{\bone}{{\mathcal B}_{1}(14\times14)}
\newcommand{\btwo}{{\mathcal B}_{2}(9\times9)}
\newcommand{\bthree}{{\mathcal B}_{3}(3\times3)}
\newcommand{\bfour}{{\mathcal B}_{4}(14\times14)}
\newcommand{\bfive}{{\mathcal B}_{5}(9\times9)}
\newcommand{\bsix}{{\mathcal B}_{6}(2\times2)}
\newcommand{\bseven}{{\mathcal B}_{7}(1\times1)}

\def\bsp#1\esp{\begin{split}#1\end{split}}
\def\bpm{\begin{pmatrix}}
	\def\epm{\end{pmatrix}}

\usepackage[normalem]{ulem}

\begin{document}

\thispagestyle{empty}
\begin{center}
\begin{Large}
\textbf{\textsc{Consistency of New CDF-II W Boson Mass with 123-Model.}}
\end{Large}

\vspace{1cm}
{
B. Ait Ouazghour,$^{1}$%
\footnote{\tt\href{mailto:brahim.aitouazghour@edu.uca.ac.ma}{brahim.aitouazghour@edu.uca.ac.ma}}
R. Benbrik$^{2}$%
\footnote{\tt \href{mailto:r.benbrik@uca.ma}{r.benbrik@uca.ma}}
E. Ghourmin$^{3}$%
\footnote{\tt \href{mailto:s.ghourmin123@gmail.com}{s.ghourmin123@gmail.com}}
M. Ouchemhou,$^{2}$%
\footnote{\tt
\href{mailto:mohamed.ouchemhou@ced.uca.ac.ma}{mohamed.ouchemhou@ced.uca.ac.ma}}
} and
L. Rahili,$^{3}$%
\footnote{\tt
\href{mailto:rahililarbi@gmail.com}{rahililarbi@gmail.com}}
\vspace*{.7cm}

{\sl $^1$ LPHEA, Faculty of Science Semlalia, Cadi Ayyad University, P.O.B. 2390 Marrakech, Morocco.}\\

{\sl
$^{2}$ Polydisciplinary Faculty, Laboratory of Fundamental and Applied Physics,
Cadi Ayyad University, Sidi Bouzid, B.P. 4162, Safi, Morocco.}\\

{\sl
$^{3}$ Laboratory of Theoretical and High Energy Physics, Faculty of Science, Ibn Zohr University, B.P 8106, Agadir, Morocco.}
\end{center}

\vspace*{0.1cm}

\begin{abstract}
Following the recent update measurement of the W boson mass performed by the CDF-II experiment at Fermilab which indicates $7\sigma$ deviation from the SM prediction. As a consequence, the open question is whether there are extensions of the SM that can carry such a remarkable deviation or what phenomenological repercussions this has. In this paper, we investigate what the theoretical constraints reveal about the \ott model. Also, we study the consistency of a CDF W boson mass measurement with the 123-model expectations, taking into account theoretical and experimental constraints. Both fit results of $S$ and $T$ parameters before and after $m_{W}^{\rm{CDF}}$ measurement are, moreover, considered in this study. Under these conditions, we found that the 123-model prediction is consistent with the measured $m_{W}^{\rm{CDF}}$ at a $95\%$ Confidence Level (CL).
\end{abstract}
 
\def\thefootnote{\arabic{footnote}}
\setcounter{page}{0}
\setcounter{footnote}{0}
\section{Introduction}
High-precision measurements at collider experiments have enacted strict constraints on the Standard Model (SM) and its possible extensions. The experimental accuracy of the electroweak observables is sensitive to the radiative corrections and needs the highest precision on the theoretical side as well. This precision measurement has been remarkably corroborated by the discovery of a Higgs particle at the LHC experiment \cite{Aad:2012tfa,Chatrchyan:2012ufa}. Moreover, it provides a pathway for deriving indirect hints on heavy new physics BSM, in particular on the not yet sufficiently explored scalar sector. In this regard, any extension of the SM is required to fulfill the $\rho \approx 1$ constraint (it is explained as being the ratio of neutral to charged current at low momentum.), which is favored by experimental data allowing just a little departure from unity.
Such a deviation, $\Delta\rho$, comes from the radiative correction taking place in the models with an extra Higgs field. $\Delta\rho$ can be related to the vector-boson self-energies and have the biggest impact in the higher-order computation of precision observables, which is a major factor to accurately measured quantities such as $m_W$, $m_Z$ as well as the effective weak mixing angle $\sin^2\theta_{\rm{eff}}$. 

\noindent
Recently, the CDF collaboration has released their newly measured $W$ boson mass \cite{CDF:2022hxs}:
\begin{align}
m^{\rm{CDF}}_W = 80.4335 \pm 0.0094\ \rm{GeV},
\end{align}  
which is out of the range of the SM prediction of about $7\sigma$, which is given by \cite{ParticleDataGroup:2020ssz,Awramik:2003rn} as follows:
\begin{align}
m^{\rm{SM}}_W = 80.357 \pm 0.006\ \rm{GeV}.
\end{align} 
Such a large deviation strongly indicates the existence of an emerging field of novel physics related to Spontaneous Symmetry Breaking (SSB), such as models with extended Higgs sectors.

\noindent
In this paper, we focus on the 123-model to investigate the possibility of predicting $m_W$ according to the new CDF measurement. The model can provide a stable candidate for dark matter and explain the tiny neutrino mass. Its scalar sector contains three CP-even Higgs bosons, $h_1$, $h_2$ and $h_3$, one CP-odd Higgs boson, $A$, a pair of charged Higgs bosons, $H^\pm$, and a pair of double-charged Higgs bosons, $H^{\pm\pm}$. Recently, several novel physics models, including the: two-Higgs doublet model \cite{Lu:2022bgw,Fan:2022dck,Zhu:2022tpr,Zhu:2022scj,Song:2022xts,Bahl:2022xzi,Heo:2022dey,Babu:2022pdn,Biekotter:2022abc,Ahn:2022xeq,Han:2022juu,Arcadi:2022dmt,Ghorbani:2022vtv, Abouabid:2022lpg, Benbrik:2022dja}, Higgs Triplet Model (HTM) \cite{Cheng:2022jyi,Du:2022brr,Kanemura:2022ahw,Mondal:2022xdy,Borah:2022obi}, Supersymmetry \cite{Yang:2022gvz,Du:2022pbp,Athron:2022isz,Zheng:2022irz,Ghoshal:2022vzo}, Leptoquark Model \cite{Athron:2022qpo,Cheung:2022zsb,Bhaskar:2022vgk}, Seesaw mechanism \cite{Blennow:2022yfm,Arias-Aragon:2022ats,Liu:2022jdq, Chowdhury:2022moc,Popov:2022ldh}, Vector-Like Leptons and/or Vector-Like Quarks \cite{Lee:2022nqz,Kawamura:2022uft,Crivellin:2022fdf,Nagao:2022oin,Chowdhury:2022dps,Cao:2022mif} and other SM extensions \cite{Strumia:2022qkt,Carpenter:2022oyg,Du:2022fqv,Yuan:2022cpw,Cacciapaglia:2022xih,Sakurai:2022hwh,Heckman:2022the,Krasnikov:2022xsi,Peli:2022ybi,FileviezPerez:2022lxp,Wilson:2022gma,Zhang:2022nnh,Alguero:2022est,Abdallah:2022shy} are proposed to explain the $W$ boson mass anomaly.

\noindent
The correction of the 123-model to $W$ boson mass can be parameterized by the $S$, $T$, and $U$ formalism as follows: \cite{Peskin:1991sw}: 
\begin{align}
\Delta m_W^2 = \big(m_W^{\rm{123}}\big)^2-\big(m_W^{\rm{SM}}\big)^2 = \frac{\alpha_0 c_W^2 m_Z^2}{c_W^2-s_W^2}\left[-\frac{1}{2}S+c_W^2 T + \frac{c_w^2-s_w^2}{4 s_W^2}U  \right],
\label{eq_mw}
\end{align}
where $\alpha_0$ is the fine structure constant at the Thomson limit, $\theta_w$ is the Weinberg angle, $m_Z$ is the Z gauge boson mass, and $s_x\,(c_x)$ stands for $\sin{\it x}\ (\cos{\it x})$.\\ 
The same formalism can, moreover, be used to study the effective weak mixing angle, $\sin^2\theta_{\rm{eff}}$, using the following relation:
\begin{align}
\Delta\sin^2\theta_{\rm{eff}} = \sin^2\theta_{\rm{eff}}\Big|_{\rm{123}}-\sin^2\theta\Big|_{\rm{SM}} = \frac{\alpha_0}{c_W^2-s_W^2} \left[ \frac{1}{4}S - s_W^2 c_W^2 T \right],
\label{eq_sintheta}
\end{align} 
wherein the SM values used are listed in Ref. \cite{ParticleDataGroup:2020ssz}. The rest of the paper is organized as follows. In Sec. \ref{sec:123model}, we describe in great length the \ott model, and then explain the theoretical investigations applied to its parameter space in Sec. \ref{subsec:theo-const}. In Sec. \ref{STU_formule}, we highlight the new physics contribution to $S$ and $T$ oblique parameters. Based on these consideration, our main results are discussed in Sec. \ref{results} and finally, we summarize our conclusions in Sec. \ref{conclusion}.
\section{The \ott model}
\label{sec:123model}
Firstly adapted in \cite{Schechter:1981cv}, the \ott model has been the focus of many studies in the past dozen years \cite{Diaz:1998zg}, the results of which provide considerable information and open up a window for new physics beyond the standard model. Nevertheless, the importance of theoretical discussions cannot be underestimated. This section sets out a brief overview of the general \ott model, including the involved multiplets, minimization conditions, and a whole discussion on Higgs bosons, gauge bosons, and neutrino mass generation in the framework of \ott mechanism.
\subsection{The scalar potential}
\label{sec:scalar-pot}
In addition to the usual SM scalar doublet, namely $\phi$, a singlet $\sigma$, and a triplet $\Delta$ have been added together to fundamentally build blocks for the \ott model. Bearing in mind its representations under the $SU(3)_{c} \times SU(2)_{L}\times U(1)_{Y}$ SM gauge group, one could explicitly write
\begin{eqnarray}
\sigma(1,1,+0) &=& \frac{1}{\sqrt{2}}(\vs+\res+i\ims),
\nonumber\\
\phi(1,2,+\frac{1}{2}) &=& \left( \begin{array}{c} \frac{1}{\sqrt{2}}(\vd+\red+i\imd) \\ \phi^- \end{array} 
\right),
\nonumber\\
\Delta(1,3,+1) &=& \left( \begin{array}{cc} \frac{1}{\sqrt{2}}(\vt+\ret+i\imt) & 
\Delta^+/\sqrt{2} \\ \Delta^+/\sqrt{2} & \Delta^{++} \end{array} \right),
\label{eq:123fields}
\end{eqnarray}
with corresponding leptonic numbers $L_{\sigma} = 2$, $L_{\phi}=0$ and $L_{\Delta}=-2$, respectively.\\ 
The most general renormalizable and gauge-invariant Lagrangian of the \ott scalar sector is given by
\begin{eqnarray}
\mathcal{L} &=&(D_\mu{\phi})^\dagger(D^\mu{\phi})+Tr(D_\mu{\Delta})^\dagger(D^\mu{\Delta})+(\partial_\mu{\sigma})^\dagger(\partial^\mu{\sigma}) -V(H, \Delta) + \mathcal{L}_{\rm Yukawa} ,
\label{eq:lag}
\end{eqnarray}
with the covariant derivatives in the kinetic terms are
\begin{eqnarray}
D_\mu \phi &=& \partial_\mu \phi + i g T^a W^a_\mu \phi + i \frac{1}{2} g^{'} B_\mu \phi, \nonumber\\
D_\mu \Delta &=& \partial_\mu \Delta + i g [T^a W^a_\mu, \Delta] + i g^{'} \frac{Y_\Delta}{2}  B_\mu \Delta,
\label{eq:cov-der}
\end{eqnarray}
where $W^a_\mu$ and $B_\mu$ stand for $SU(2)_L$ and $U(1)_Y$ gauge fields, respectively. $Y_\Delta$ is the hypercharge operator of the triplet $\Delta$, while $T^a$ is related to the Pauli matrices via $T^a=\sigma^a/2$. $\mathcal{L}_{\rm Yukawa}$ refer to the Yukawa part to be subsequently considered in detail. 

\noindent
The scalar potential $V(\sigma,\phi,\Delta)$, invariant under $SU(2)_L \times U(1)_Y$, reads \cite{Akeroyd:2010eg,Blunier:2016peh}  
\begin{align}
\VHiggs &= \musig^2 \sigma^\dagger \sigma + \muphi^2 \phi^\dagger \phi
+ \mudel^2 {\mathrm{Tr}} ( \Delta^\dagger \Delta ) + \lambda_1 (\phi^\dagger \phi)^2\nonumber \\
&+\lambda_2 \big[{\mathrm{Tr}} ( \Delta^\dagger \Delta )\big]^2 +\lambda_3 (\phi^\dagger \phi) {\mathrm{Tr}} ( \Delta^\dagger \Delta )
\nonumber\\ 
&+ \lambda_4 {\mathrm{Tr}} ( \Delta^\dagger \Delta \Delta^\dagger \Delta )
+ \lambda_5 ( \phi^\dagger \Delta^\dagger \Delta \phi ) + \beta_1 (\sigma^\dagger \sigma)^2\nonumber \\
&+ \beta_2 (\phi^\dagger \phi) (\sigma^\dagger \sigma)
+ \beta_3 {\mathrm{Tr}} ( \Delta^\dagger \Delta ) (\sigma^\dagger \sigma)
\nonumber\\ &
-\kappa (\phi^T \Delta \phi \sigma + \text{h.c.}),
\label{eq:sc-pot}
\end{align}
where all quartic couplings are considered to be real. $\mu^2_i$ ($i=\sigma, \phi, \Delta$) are squared mass parameters of the singlet, doublet, and triplet fields, respectively. These parameters can be eliminated by imposing the following vacuum conditions:
\begin{align}
2 \vs \musig^2 &= \kappa \vt \vd^2 - 2\beta_1 \vs^3 - \beta_2 \vs \vd^2 -  \beta_3 \vs \vt^2, \nonumber \\
2 \muphi^2 &= 2\kappa \vs \vt - \beta_2 \vs^2 - 2\lambda_1 \vd^2 - \lambda_3 \vt^2 - \lambda_5 \vt^2, \nonumber \\
2 \vt \mudel^2 &= \kappa \vs \vd^2 - \beta_3 \vs^2 \vt -2\lambda_2 \vt^3 - \lambda_3 \vt \vd^2 -  2\lambda_4 \vt^3 - \lambda_5 \vt \vd^2,
\label{eq:vac-cond}
\end{align}
thereby reducing the set of free parameters down by three degree-of-freedom.

\subsection{The ﬁeld composition of the model}
\label{sec:ﬁeld-composition}
The next stage is to extract from the Lagrangian basis the usual mass matrices for the \ott model. The bilinear part of the Higgs potential in Eq.(\ref{eq:sc-pot}) is given by
\begin{align}
\VHiggs &\supset \begin{pmatrix} \delta^{--} \end{pmatrix} \matr{M}_{\delta^{\pm\pm}}^2
\begin{pmatrix} \delta^{++} \end{pmatrix} + \begin{pmatrix} \phi^-,\delta^-  \end{pmatrix}
\matr{M}_{\phi^\pm\delta^\pm}^2 \begin{pmatrix} \phi^+ \\ \delta^+  \end{pmatrix} + \frac{1}{2}\begin{pmatrix} \res,\red, \ret \end{pmatrix} \matr{M}_{R}^2 \begin{pmatrix} \res \\ \red \\ \ret \end{pmatrix} \nonumber\\
&+\frac{1}{2} \begin{pmatrix} \ims, \imd, \imt \end{pmatrix} \matr{M}_{I}^2 \begin{pmatrix} \ims \\ \imd \\ \imt \end{pmatrix} + \cdots
\end{align}
where $\matr{M}_{\delta^{\pm\pm}}^2$, $\matr{M}_{\phi^\pm\delta^\pm}^2$, $\matr{M}_{R}^2$ and $\matr{M}_{I}^2$ are $1\times1$, $2\times2$, $3\times3$ and $3\times3$ mass matrices of the doubly charged, simply charged, CP-even sector and CP-odd sector, respectively. 

At tree-level and without any assumption except Eq.(\ref{eq:vac-cond}), the three entries of the scalar and pseudo-scalar Higgs sector matrices are given by \cite{Blunier:2016peh}
\begin{equation}
\begin{aligned}
\left(\matr{M}_{R}\right)_{11}^{2} &= 2\beta_1\vs^2 + \frac{1}{2}\kappa \vd^2 \frac{\vt}{\vs} ,  &        
\left(\matr{M}_{I}\right)_{11}^{2} &= \frac{1}{2} \kappa \vd^2 \frac{\vt}{\vs},   \\
\left(\matr{M}_{R}\right)_{22}^{2} &= 2 \lambda_1 \vd^2 ,  &                  
\left(\matr{M}_{I}\right)_{22}^{2} &= 2 \kappa \vt \vs, \\
\left(\matr{M}_{R}\right)_{33}^{2} &= 2(\lambda_2+\lambda_4)\vt^2 +\frac{1}{2} \kappa \vd^2 \frac{\vs}{\vt} ,  &                  
\left(\matr{M}_{I}\right)_{33}^{2} &= \frac{1}{2} \kappa \vd^2 \frac{\vs}{\vt}, \\
\left(\matr{M}_{R}\right)_{12}^{2} &= \beta_2 \vd \vs-\kappa \vd \vt = \left(\matr{M}_{R}\right)_{21}^{2} ,  &                  
\left(\matr{M}_{I}\right)_{12}^{2} &= \kappa \vd \vt = \left(\matr{M}_{I}\right)_{21}^{2} ,\\
\left(\matr{M}_{R}\right)_{13}^{2} &=  \beta_3 \vt \vs-\frac{1}{2}\kappa \vd^2 = \left(\matr{M}_{R}\right)_{31}^{2} ,   &               
\left(\matr{M}_{I}\right)_{13}^{2} &=  \frac{1}{2}\kappa \vd^2 = \left(\matr{M}_{I}\right)_{31}^{2} ,\\
\left(\matr{M}_{R}\right)_{23}^{2} &=  (\lambda_3+\lambda_5)\vd \vt-\kappa \vd \vs = \left(\matr{M}_{R}\right)_{32}^{2},  &                  
\left(\matr{M}_{I}\right)_{23}^{2} &=  \kappa \vd \vs = \left(\matr{M}_{I}\right)_{32}^{2},\\
\end{aligned}
\end{equation}
while for the charged sectors read as \cite{Blunier:2016peh},
\begin{equation}
\begin{aligned}
\left(\matr{M}_{\delta^{\pm\pm}}^{2}\right) &= - \lambda_4 \vt^2 - \frac{1}{2} \lambda_5 \vd^2 + \frac{1}{2} \kappa \vd^2 \frac{\vs}{\vt} , \\
\left(\matr{M}_{\phi^\pm\delta^\pm}\right)_{11}^{2} &= -\frac{1}{2} \lambda_5 \vt^2 + \kappa \vt \vs , \\
\left(\matr{M}_{\phi^\pm\delta^\pm}\right)_{22}^{2} &= -\frac{1}{4} \lambda_5 \vd^2 + \frac{1}{2} \kappa \vd^2 \vs / \vt , \\
\left(\matr{M}_{\phi^\pm\delta^\pm}\right)_{12}^{2} &= \frac{1}{2\sqrt{2}}\lambda_5\vt \vd - \frac{1}{\sqrt{2}} \kappa \vd \vs = \left(\matr{M}_{\phi^\pm\delta^\pm}\right)_{21}^{2}, \\
\end{aligned}
\end{equation}
By applying a unitary transformation in the non-physical fields basis, one can get the mass eigenstates in the lowest order as follows:
\begin{align}
\begin{pmatrix} h_1 \\ h_2 \\ h_3 \end{pmatrix} 
= \Ueven \cdot
\begin{pmatrix} \res \\ \red \\ \ret \end{pmatrix}, 
\quad
\begin{pmatrix} G \\ J \\ A \end{pmatrix} 
= \Uodd \cdot
\begin{pmatrix} \ims \\ \imd \\ \imt \end{pmatrix}
\quad{\rm and}\quad
\begin{pmatrix} G^\pm \\ H^\pm \end{pmatrix} 
= \UChar \cdot
\begin{pmatrix} \phi^\pm \\ \Delta^\pm \end{pmatrix}.
\label{eq:higgs-sc-rotation}
\end{align}
The two $3\times3$ matrices $\Ueven$ and $\Uodd$ transform the neutral $\cp$-even and $\cp$-odd Higgs fields,
respectively. They take the following form \cite{ParticleDataGroup:2016lqr, Blunier:2016peh}:
\beq
\Ueven &=&\left( \begin{array}{ccc}
	c_{\alpha_1} c_{\alpha_2} & s_{\alpha_1} c_{\alpha_2} & s_{\alpha_2}\\
	-(c_{\alpha_1} s_{\alpha_2} s_{\alpha_3} + s_{\alpha_1} c_{\alpha_3})
	& c_{\alpha_1} c_{\alpha_3} - s_{\alpha_1} s_{\alpha_2} s_{\alpha_3}
	& c_{\alpha_2} s_{\alpha_3} \\
	- c_{\alpha_1} s_{\alpha_2} c_{\alpha_3} + s_{\alpha_1} s_{\alpha_3} &
	-(c_{\alpha_1} s_{\alpha_3} + s_{\alpha_1} s_{\alpha_2} c_{\alpha_3})
	& c_{\alpha_2}  c_{\alpha_3}
\end{array} \right) = V_{\rm PMNS},
\label{eq:mixingmatrix1}\\
\Uodd &=&\left( \begin{array}{ccc}
	0 & \frac{1}{N_G} &  -\frac{2}{N_G} \frac{\vt}{\vd}  \\
	\frac{N_G^2}{N_J} & -\frac{2}{N_J} \frac{\vt^2}{\vd\vs} &  -\frac{1}{N_J} \frac{\vt}{\vs} \\
	\frac{1}{N_A} \frac{\vt}{\vs} & \frac{2}{N_A} \frac{\vt}{\vd} &  \frac{1}{N_A}
\end{array} \right),
\label{eq:mixingmatrix2}
\eeq
The mixing angles $\alpha_{1,2,3}$ could lie within the range $[-\pi/2, +\pi/2]$, and $N_{G,J,A}$ are defined as follows
\begin{equation}
N_G = \sqrt{ 1 + 4 \frac{v_\Delta^2}{v_\phi^2} },\,\, N_J = \sqrt{ N^4_G + 4 \frac{v_\Delta^4}{v_\phi^2 v_\sigma^2} + \frac{v_\Delta^2}{v_\sigma^2}}\,\,\,{\rm and}\,\,\, N_A = \sqrt{1+4\frac{v_\Delta^2}{v_\phi^2}+\frac{v_\Delta^2}{v_\sigma^2}},
\end{equation}
whereas $\UChar$ is the matrix that transforms the charged Higgs field which is given by its $2\times2$ representation $\{\{-c_\beta,s_\beta\},\{s_\beta,c_\beta\}\}$.  $s_\beta\ (c_\beta)$ stands for $\sin\beta\ (\cos\beta)$ satisfying $t_\beta=\tan\beta=\sqrt{2}\vt/\vd$.

Based on the foregoing, the Higgs sector of \ott model is made up of nine scalar bosons, five of them being electrically neutral (denoted usually as $h_1$, $h_2$, $h_3$, $A$ and the Majoron $J$) and the other four charged ($H^{\pm}$ and $H^{\pm\pm}$). Their masses can be written as
\begin{eqnarray}
\text{diag}(m^2_{h_1},m^2_{h_2},m^2_{h_3}) & = & \Ueven \matr{M}_{R} {\Ueven}^{\text T}, \\
m_A^2 & = & \frac{1}{2} \kappa \bigg( \frac{v_\sigma v_\phi^2}{v_\Delta} + \frac{v_\Delta v_\phi^2}{v_\sigma}
+ 4 v_\sigma v_\Delta \bigg), \\
m_J^2 & = & 0, \\
m_{H^\pm}^2 & = & \frac{1}{2} \bigg( \kappa\frac{v_\sigma}{v_\Delta} - \frac{1}{2} \lambda_5 \bigg)
\Big( v_\phi^2 + 2 v_\Delta^2 \Big),\\
m_{H^{\pm\pm}}^2 & = & - \lambda_4 v_\Delta^2 - \frac{1}{2} \lambda_5 v_\phi^2 +
\frac{1}{2} \kappa v_\phi^2 \frac{v_\sigma}{v_\Delta}.
\end{eqnarray}
Here the Majoron $J$ is the second massless physical Higgs in the $\cp$-odd sector to be predominantly singlet, 
matching the consistency of the \ott model with the LEP measurements of the invisible $Z$ decay width \cite{ParticleDataGroup:2014cgo,Carena:2003aj}. Furthermore, it is worth to mention that a different hierarchy between $H^{\pm \pm}$, $H^{\pm}$ and $A$ masses can occur, and mainly depends on $\lambda_5$ sign, resulting in splitting that is described by  (assuming $\vt \ll \vd$)
\begin{equation}
\Delta m^2 \approx m_{H^{\pm\pm}}^2-m_{H^{\pm}}^2 \approx m_{H^{\pm}}^2-m_{A}^2 \approx \frac{1}{2}(m_{H^{\pm\pm}}^2-m_{A}^2).
\label{diffchdmass}
\end{equation}
%


\subsection{The Model Parameters}
Let's begin by setting two redefinitions of VEV's as follows:
\beq
v^2 = \vd^2 + 2\vt^2 \quad {\rm and} \quad v_0^4 = 4\vs^2\vt^2 + \vd^2(\vs^2+\vt^2),
\eeq
in terms of the physical basis parameters, the dimensionless quartic couplings, $\lambda_i$, of the \ott  potential, which read 
\beq
\kappa &=& \frac{2 \vs\vt}{v_0^4} m_A^2 \label{eq:kap}\\
\lambda_1 &=& \frac{1}{2 \vd^2} \sum_{i=1}^3 (\Ueven_{i2})^2 m_{h_i}^2 \label{eq:la1} \\
\lambda_2 &=& \frac{1}{2 \vt^2} \left( \sum_{i=1}^3 (\Ueven_{i3})^2 m_{h_i}^2 - 4 \frac{\vd^2}{v^2} m_{H^\pm}^2 + 2 m_{H^{\pm\pm}}^2 +\frac{\vd^2\vs^2}{v_0^4} m_A^2 \right) \label{eq:la2}\\
\lambda_3 &=& \frac{1}{\vd \vt}  \sum_{i=1}^3 \Ueven_{i2}\Ueven_{i3} m_{h_i}^2 + \frac{4}{v^2} m_{H^\pm}^2 - \frac{2\vs^2}{v_0^4} m_A^2  \label{eq:la3}\\
\lambda_4 &=& \frac{1}{\vt^2} \left(2 \frac{\vd^2}{v^2} m_{H^\pm}^2 - m_{H^{\pm\pm}}^2 - \frac{\vs^2\vd^2}{v_0^4} m_A^2 \right) \label{eq:la4}\\
\lambda_5 &=& 4 \left(\frac{\vs^2}{v_0^4} m_A^2 - \frac{1}{v^2} m_{H^\pm}^2 \right) \label{eq:la5}\\
\beta_1 &=& \frac{1}{2 \vs^2} \left( \sum_{i=1}^3 (\Ueven_{i1})^2 m_{h_i}^2 - \frac{\vd^2\vt^2}{v_0^4} m_A^2 \right) \label{eq:bt1}\\
\beta_2 &=& \frac{1}{\vs\vd} \left( \sum_{i=1}^3 \Ueven_{i1}\Ueven_{i2} m_{h_i}^2 + \frac{2\vs\vd\vt^2}{v_0^4} m_A^2 \right) \label{eq:bt2}\\
\beta_3 &=& \frac{1}{\vs\vt} \left( \sum_{i=1}^3 \Ueven_{i1}\Ueven_{i3} m_{h_i}^2 + \frac{\vs\vd^2\vt}{v_0^4} m_A^2 \right) \label{eq:bt3}
\eeq

To end with this part, the \ott model, in total, is described by 12 independent real degrees of freedom. By considering the further constraint $v = \sqrt{\vd^2+ 4\vt^2}$ arising from the correct electroweak scale requirements, and using the previous Eq.(\ref{eq:vac-cond}) to trade the three multiplet masses for the SM Vacuum Expectation Value (VEV) $v$, $t_\beta$ and $v_S$. Thus, we use the following hybrid set of input parameters:
\beq
m_{h_{1}} \,,\,\, m_{A} \,,\,\, \lambda_j\,(i=2,4,5) \,,\,\, m_{H^{\pm}} \,,\,\, \beta_3 \,,\,\, \alpha_{i\,(i=1,2,3)} \,,\,\, \vs \,,\,\, \vt. \label{eq:n2hdminputpars}
\eeq

\section{Theoretical Constraints on Lagrangian Parameters}
\label{subsec:theo-const}
Before proceeding with a complete scan over the whole space parameters, it may be recalled that the \ott model has been and continues to be a matter of many theoretical and experimental investigations. The first one relates mainly to : boundedness from below of the scalar potential, perturbative unitarity, and the global minimum that the potential must preserve.    

\subsection{Vacuum Stability}
\label{subsec:vacuum-sta}
A prerequisite was to ensure that scalar potential of the \ott model is bounded from below, where the quartic terms assert itself at large field strength ($V>-\infty$). Following the same methodology as in \cite{Arhrib:2011uy,Bonilla:2015eha,Ouazghour:2018mld,Arhrib:2018qmw}, the authors in \cite{Pinheiro:2020zde} have derived the constraints ensuring BFB and the corresponding whole set read :
\begin{align}
\lambda_1 > 0 & , & \beta_1 > 0 &, \nonumber\\
\lambda_2 + \lambda_4 > 0 & , & \beta_2 + 2\kappa + 2 \sqrt{\beta_1 \lambda_1} > 0 &, \nonumber\\
\lambda_2 + \lambda_4/2 > 0 & , & \beta_2 - 2\kappa + 2 \sqrt{\beta_1 \lambda_1} > 0 &, \nonumber\\
\lambda_3 + 2 \kappa  + 2 \sqrt{\lambda_1(\lambda_2 + \lambda_4)} > 0 & , & \beta_3 + 2\kappa + 2 \sqrt{\beta_1 (\lambda_2 + \lambda_4)} > 0 &, \nonumber\\
\lambda_3 + 2 \kappa  + 2 \sqrt{\lambda_1(\lambda_2 + \lambda_4/2)} > 0 & , & \beta_3 + 2\kappa + 2 \sqrt{\beta_1 (\lambda_2 +  \lambda_4/2)} > 0 &, \nonumber\\
\lambda_3 + \lambda_5 + 2 \kappa + 2 \sqrt{\lambda_1(\lambda_2 + \lambda_4/2)} > 0 & , & \beta_3 - 2\kappa + 2 \sqrt{\beta_1 (\lambda_2 + \lambda_4/2)} > 0 &, \nonumber\\
\lambda_3 + \lambda_5 + 2 \kappa + 2 \sqrt{\lambda_1(\lambda_2 +  \lambda_4)} > 0 & , & \beta_3 - 2\kappa + 2 \sqrt{\beta_1 (\lambda_2 + \lambda_4/2)} > 0 &, \nonumber\\
\lambda_3 - 2 \kappa  + 2 \sqrt{\lambda_1(\lambda_2 + \lambda_4/2)} > 0 & , & \beta_3 - 2\kappa + 2 \sqrt{\beta_1 (\lambda_2 +  \lambda_4)} > 0 &, \nonumber\\
\lambda_3 - 2 \kappa  + 2 \sqrt{\lambda_1(\lambda_2 + \lambda_4)} > 0 & , &  & \nonumber\\
\lambda_3 + \lambda_5 - 2 \kappa + 2 \sqrt{\lambda_1(\lambda_2 + \lambda_4/2)} > 0 & , &  & \nonumber\\
\lambda_3 + \lambda_5 - 2 \kappa + 2 \sqrt{\lambda_1(\lambda_2 + \lambda_4)} > 0 & , &  &
\end{align}
%

\subsection{$\mathcal{S}-$Matrix unitarity}
\label{subsec:unita}
A closer look reveals that the entire set of $2$-body scalar scattering processes results in a $52\times52$ $S$-matrix that can be split up into 7 block submatrices representing mutually unmixed groups of channels with definite charge and CP states, organized in a database $\mathcal{B}_{i}$ in terms of net electric charge in the initial/final states:   
$\bone$, $\btwo$ and $\bthree$, corresponding to 
$0$-charge channels, $\bfour$ corresponding to the $1$-charge channels, 
$\bfive$ corresponding to the $2$-charge channels, $\bsix$ corresponding to the $3$-charge channels, and finally $\bseven$ corresponding to the $4$-charge channels. The following Table \ref{tab:basis-eigenvalues-1} highlights an illustration of the framework described above.  
\begin{table}[!h]
	\centering
	\begin{tabular}{cll}
		\hline \hline
		$Q$ & Basis states &  Eigenvalues  \\
		\hline
		0  &  $\mathcal{B}_{1}=\big\{\phi^+\delta^-,\,\delta^+\phi^-,\,\red\imt,\,\red\ims,\,\ret\imd,\,\ret\ims,\,\res\imd,$  &   $a_{1}^+,\, a_{1}^-,\, b_{1},\, b_{2},\, b_{3}$ \\
		&  \hspace{1cm} $\res\imt,\,\res\ret,\,\res\red,\,\red\ret,\,\ims\imt,\,\ims\imd,\,\imd\imt\big\}$  &  $b_{4}$   \\ 
		\hline  
		0  &  $\mathcal{B}_{2}=\big\{\phi^+\phi^-,\,\delta^+\delta^-,\,\delta^{++}\delta^{--},\,R_\phi R_\phi/\sqrt{2},\,R_\Delta R_\Delta/\sqrt{2},$  &   $a_{2}^+,\, a_{2}^-,\, b_{5},\, b_{6},\, b_{7}$ \\
		&  \hspace{1cm} $R_\sigma R_\sigma/\sqrt{2},\,I_\phi I_\phi/\sqrt{2},\,I_\Delta I_\Delta/\sqrt{2},\,I_\sigma I_\sigma/\sqrt{2}\big\}$  &  $b_{8},\, b_{9...11}$ \\
		\hline  
		0  &  $\mathcal{B}_{3}=\big\{R_\phi I_\phi,\,R_\Delta I_\Delta,\,R_\sigma I_\sigma\big\}$  &  $b_{5},\, b_{6},\, b_{8}$ \\
		\hline
		1  &  $\mathcal{B}_{4}=\big\{R_\phi\phi^+,\,R_\Delta\phi^+,\,R_\sigma\phi^+,\,I_\phi\phi^+,\,I_\Delta\phi^+,\,I_\sigma\phi^+,\,R_\phi\delta^+,$  &   $a_{1}^+,\, a_{1}^-,\, a_{2}^+,\, a_{2}^-,\, a_{3}^+,\, a_{3}^-$  \\
		&  \hspace{1cm} $R_\Delta\delta^+,\,R_\sigma\delta^+,\,I_\phi\delta^+,\,I_\Delta\delta^+,\,I_\sigma\delta^+,\,\delta^{++}\phi^{-},\,\delta^{++}\delta^{-}\big\}$  &  $b_{1},\, b_{2},\, b_{3},\, b_{4},\, b_{7},\, b_{8},\, b_{12}$   \\
		\hline
		2  &  $\mathcal{B}_{5}=\big\{\phi^+\phi^+/\sqrt{2},\,\delta^+\delta^+/\sqrt{2},\,\phi^+\delta^{+},\,\delta^{++}R_\phi,\,\delta^{++}R_\Delta,$  &   $a_{3}^+,\, a_{3}^-,\, b_{2},\, b_{3},\, b_{4}$ \\
		&  \hspace{1cm} $\delta^{++}R_\sigma,\,\delta^{++}I_\phi,\,\delta^{++}I_\Delta,\,\delta^{++}I_\sigma\big\}$  &  $b_{7},\,b_{8},\, b_{12},\, b_{13}$ \\  
		\hline
		3 & $\mathcal{B}_{6}=\big\{\delta^{++}\phi^+,\,\delta^{++}\delta^{+}\big\}$ & $b_3,\,b_8$  \\
		\hline
		4 & $\mathcal{B}_{7}=\big\{\delta^{++}\delta^{++}/\sqrt{2}\big\}$ & $b_{8}$  \\
		\hline \hline
	\end{tabular}
	\caption{Basis states and eigenvalues of the complete set of $2$-body scalar scattering processes matrix $\mathcal{M}$, categorized based on the overall charge $Q$ of the initial and final states.}
	\label{tab:basis-eigenvalues-1}
\end{table}

The complete set of eigenvalues electromagnetism, described as a combinations of $\lambda'$s couplings are given by :
\begin{align}
b_1 &= \beta_2, &
a_1^{\pm}  &= \frac{1}{4} \Big(2\beta_2 + 2\lambda_3 + 3\lambda_5 \pm \sqrt{\left(2\beta_2-2\lambda_3-3\lambda_5\right)^2 + 96\kappa^2} \, \Big) \nonumber\\
b_2 &= \lambda_3, &
a_2^{\pm}  &= \lambda_1 + \lambda_2 + 2 \lambda_4 \pm \sqrt{\left(\lambda_1 - \lambda_2 - 2 \lambda_4\right)^2 + \lambda_5^2}, \nonumber\\
b_3 &= \lambda_3 + \lambda_5, &
a_3^{\pm}  &= \frac{1}{2} \Big(\beta_3 + 2\lambda_1 \pm \sqrt{\left(\beta_3 - 2\lambda_1\right)^2 + 8\kappa^2} \, \Big) , \nonumber\\
b_4 &= \beta_3, &
b_8  &= 2 \Big(\lambda_2 + \lambda_4\Big), \nonumber\\
b_5 &= 2\beta_1, &
b_{12}  &= \lambda_3 -\frac{1}{2} \lambda_5, \nonumber\\
b_6 &= 2\lambda_1, &
b_{13}  &= 2\lambda_2-\lambda_4, \nonumber\\
b_{7} &= 2\lambda_2. &
\label{eq:lam_coup}
\end{align}

\noindent
whereas the remaining $b_{9...11}$ eigenvalues are the roots of a third-order equation given by
\begin{eqnarray}
&&x^3 
- 8 \Big[ 3 \lambda_1 + 3 \lambda_4 + 4 \lambda_2 + 2 \beta_1 \Big] x^2
+ 8 \Big[ 8 \big( 3 \lambda_1 (4 \lambda_2 + 3 \lambda_4) + \beta_1 (6 \lambda_1 + 8 \lambda_2 + 6 \lambda_4) \big) - \nonumber\\
&& 3 (2 \lambda_3 + \lambda_5)^2 - 4 \beta_2^2 - 6 \beta_3^2 \Big] x
-128 \Big[-9\lambda_1\beta_3^2 - 2 (4 \lambda_2 + 3 \lambda_4 ) \beta_2^2 + 3 ( 2 \lambda_3 + \lambda_5 ) \beta_2 \beta_3 + \nonumber\\
&& 3 \beta_1 \big( 8 \lambda_1 (4 \lambda_2 + 3 \lambda_4) - (2 \lambda_3 + \lambda_5)^2 \big) \Big]
\end{eqnarray}
%

\subsection{Electroweak minimum}
\label{subsec:elec-min}
By use of the three minimization equations in \ref{eq:vac-cond}, one should therefore find a configuration from all the $(\sigma,\phi,\Delta)$ space values such that the scalar potential is in a minimum situation. For such purpose, we redefine the fields in equation \ref{eq:123fields} by assuming that electroweak symmetry breaking is taking place; in this way, the structure of the potential would be energetically favored for $\langle V \rangle_{\rm EWSB} <0$. Thereby, the naive bound on $\kappa$
\begin{eqnarray}
\kappa < \kappa_{\rm max} \equiv \frac{\vs^3}{2\vd^2\vt}\beta_1+\frac{\vs}{2\vt}\beta_2+\frac{\vs\vt}{2\vd^2}\beta_3+\frac{\vd^2}{2\vs\vt}\lambda_1+
\frac{\vt^3}{2\vd^2\vs}\lambda_{24}^++
\frac{\vt}{2\vs}\lambda_{35}^+  
\end{eqnarray}
where $\lambda_{ij}^+=\lambda_i+\lambda_j$, is a necessary and sufficient one to ensure $\langle V \rangle_{\rm EWSB} <0$, and hence the minimum to be unique.

\section{Bounds from the Electroweak Precision tests}
\label{STU_formule}
To provide high precision for the \ott-space parameter as an electroweak theory, additional precision has to be studied, which could impose severe bounds on the new physics. Accordingly, we highlight the Peskin-Takeuchi parameters $S$, $T$, and $U$ \cite{Peskin:1991sw} in the \ott model. Where the $S$ parameter measures the deviation from the SM prediction for the electroweak (EW) radiative correction, which describes the breaking of the weak isospin symmetry, however, the $T$ parameter measures the deviation from the SM prediction for the weak isospin symmetry breaking in the heavy sector, which is related to the difference between the masses of the W and $Z$ bosons. In the \ott model with $v_t=1$ or less, there is a decoupling between the doublet field and singlet field on one side and the triplet field on the other side. This decoupling is done only in the doubly charged, simply charged, and CP-odd sectors. That is to say that the major contribution of physical fields $H^\pm$, $A$ and $H^{\pm\pm}$ comes from triplet fields, and the major contribution of physical fields $h_1$, $h_2$, $h_3$ and $J$ comes from doublet and singlet fields. We use the general expressions presented in \cite{Ghosh:2022cca,Lavoura:1993nq,grimus2008precision,grimus2008oblique}. Approximative contributions of new scalar fields to $S$ and $T$ parameters in the \ott model are then given by:
\begin{eqnarray}
S &=& \frac{1}{24 \pi}\bigg[\big({U_{11}^H}^2+{U_{12}^H}^2+4{U_{13}^H}^2\big)\ln{m_{h_1}^2}+\big({U_{21}^H}^2+{U_{22}^H}^2+4{U_{23}^H}^2\big)\ln{m_{h_2}^2}\nonumber\\
&+&2\big({U_{31}^H}^2+{U_{32}^H}^2+4{U_{33}^H}^2\big)\ln{m_{h_3}^2}-\ln{m_{h_{\rm ref}}^2}+{{\cal U}_{11}^H}^2 \hat G(m_{h_1}^2,m_{Z}^2)+{{\cal U}_{12}^H}^2 \hat G(m_{h_2}^2,m_{Z}^2)\nonumber\\
&+&2 G(m_{h_3}^2,m_{h_3}^2,m_{Z}^2)-\hat G(m_{h_{\rm ref}}^2,m_{Z}^2)\bigg]-\frac{1}{3\pi}\bigg[\ln\left(m_{H^{\pm\pm}}^2\right)-\frac{(1-2s^2_w)^2}{2}\xi\left(m_{H^{\pm\pm}}^2,m_{H^{\pm\pm}}^2,m_Z^2\right)\nonumber\\
&-&\frac{s^4_w}{2} \xi\left(m_{H^{\pm}}^2,m_{H^{\pm}}^2,m_Z^2\right)\bigg]	
\label{S}, 
\end{eqnarray}
and 
\begin{eqnarray}
T &=& \frac{1}{16 \pi m_W^2 s_W^2}\bigg[ F(m_{H^{\pm\pm}}^2,m_{H^\pm}^2)+{U_{12}^H}^2\,F(m_{H^\pm}^2,m_{h_1}^2)+{U_{22}^H}^2\,F(m_{H^\pm}^2,m_{h_2}^2)\nonumber	\\
&+&{U_{32}^H}^2\,F(m_{H^\pm}^2,m_{h_3}^2)+3\,{{\cal U}_{11}^H}^2\,\big(F(m_Z^2,m_{h_1}^2)-F(m_W^2,m_{h_1}^2)\big)+3\,{{\cal U}_{12}^H}^2\,\big(F(m_Z^2,m_{h_2}^2)	\\
&-&F(m_W^2,m_{h_2}^2)\big)+3\,{{\cal U}_{13}^H}^2\,\big(F(m_Z^2,m_{h_3}^2)-F(m_W^2,m_{h_3}^3)\big)-3\,\big(F(m_Z^2,m_{h_{\rm ref}}^2)-F(m_W^2,m_{h_{\rm ref}}^2)\big)  \bigg] \nonumber
\end{eqnarray}
with
\begin{equation}
{\cal U}_{11}^H = U^A_{11}U^H_{11}+U^A_{12}U^H_{12}+2U^A_{13}U^H_{13}
\end{equation} 
\begin{equation}
{\cal U}_{12}^H = U^A_{11}U^H_{21}+U^A_{12}U^H_{22}+2U^A_{13}U^H_{23}
\end{equation} 
\begin{equation}
{\cal U}_{13}^H = U^A_{11}U^H_{31}+U^A_{12}U^H_{32}+2U^A_{13}U^H_{33}
\end{equation}
whilst $m_{h_{ref}}$ stands for the SM reference represented by $m_{h_{ref}}=125.09$ GeV, and the functions $F$, $\xi$ and $\hat G$ can be found in Refs \cite{Ghosh:2022cca,Lavoura:1993nq}.
Note that, the precise measurements of the electroweak precision observables, such as the W and $Z$ boson masses and the electroweak mixing angle, are used to determine the values of the $S$ and $T$ parameters. The experimental values of the $S$ and $T$ parameters can be used to constrain models of new physics.
\section{Results and discussion}
\label{results}
In order to examine whether the CDF $m_W$ mass measurement is consistent with the \ott model's theoretical framework, we randomly scan over its parameter space as indicated in Table \ref{tab:tab1}.
\begin{table}[!h]
\begin{center}
\begin{tabular}{ l | c l}
\hline\hline    
$m_{h_{1}}$     & &   125 \\  
$\lambda_{2}$             & &  [-6,8$\pi$]\\ 
$\lambda_{4}$             & &  [-9,8$\pi$]\\ 
$\lambda_{5}$             & &  [-15,14]\\ 
$\beta_3$               & &  [-8$\pi$,8$\pi$] \\
$\kappa$             & &  [0,0.1]\\
$v_{\Delta}$             & &  [0,1] \\
$v_{s}$            & &  [10,1000] \\
$\alpha_{1}$             & &  [-$\pi$/2,$\pi$/2] \\
$\alpha_{2}$             & &  [-0.5,0.5] \\
$\alpha_{3}$             & &  [-$\pi$/2,$\pi$/2] \\
\hline\hline    
\end{tabular}
\end{center}
\caption{\ott parameter space scan (all masses and {\it vev}'s are in GeV).}
\label{tab:tab1}
\end{table}

\noindent
As previously mentioned, we assume that the \cp-even $h_1$ plays the role of the SM-like Higgs boson with a mass near $125$ GeV, which characteristics match the LHC measurements. In addition, to obtain a viable model, we require all \ott parameter points to satisfy the following theoretical and experimental constraints:
\begin{itemize}
\item Unitarity\footnote{We notice that such constraints were generated for the first time within this framework.}, perturbativity, and vacuum stability requirements.
\end{itemize} 
\begin{itemize}
\item The consistency with the $95\%$ Bounds imposed by the LHC are checked via the public program HiggsBounds-5.10.2 \cite{Bechtle:2020pkv}.
\end{itemize} 
\begin{itemize}
\item The criterion that the \cp-even $h_1$ Higgs boson need to match the characteristics of the detected SM-like Higgs boson is enforced using the public code HiggsSignal-2.6.2 \cite{Bechtle:2020uwn}. 
\end{itemize} 
\begin{figure}[!hb]
\centering
\hspace*{0.5cm}
\includegraphics[width=0.45\textwidth]{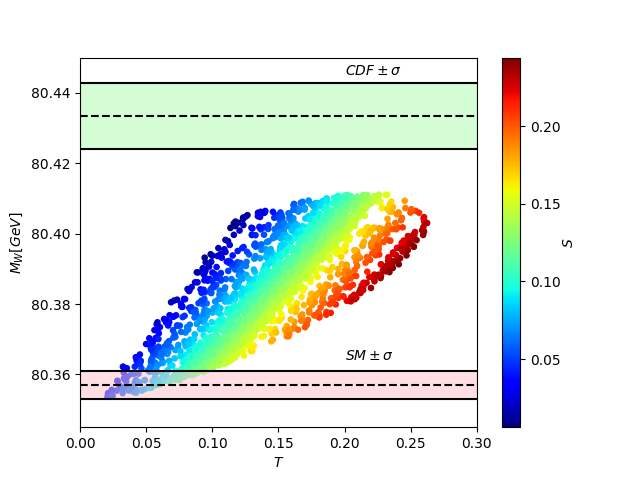}%
\includegraphics[width=0.45\linewidth]{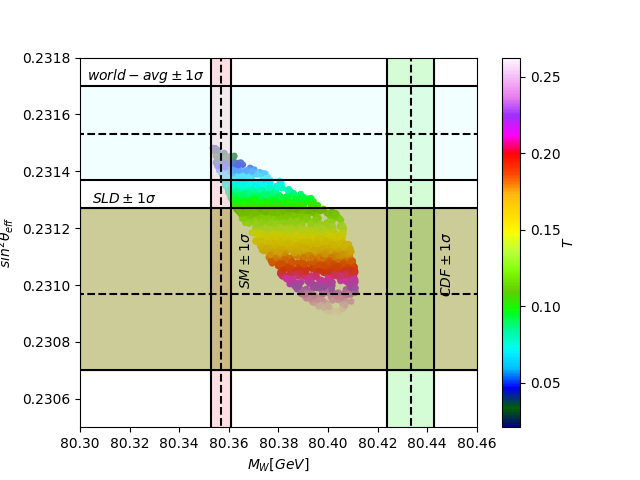}
\\
\hspace*{0.25\textwidth} 
(a) \hfill (b)
\hspace*{0.25\textwidth} 
\\
\hspace*{0.5cm}
\includegraphics[width=0.45\linewidth]{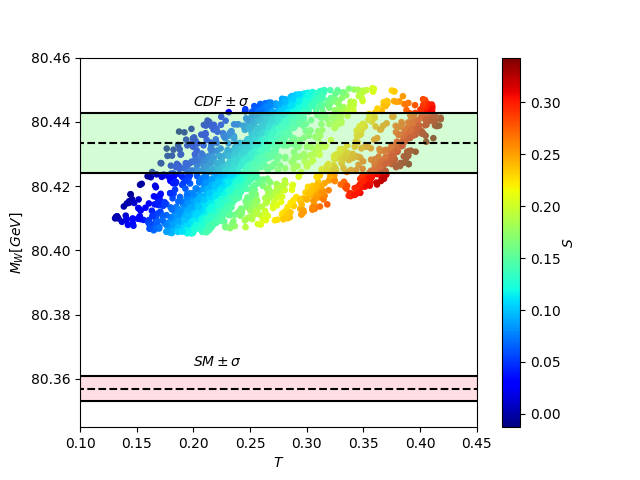}%
\includegraphics[width=0.45\linewidth]{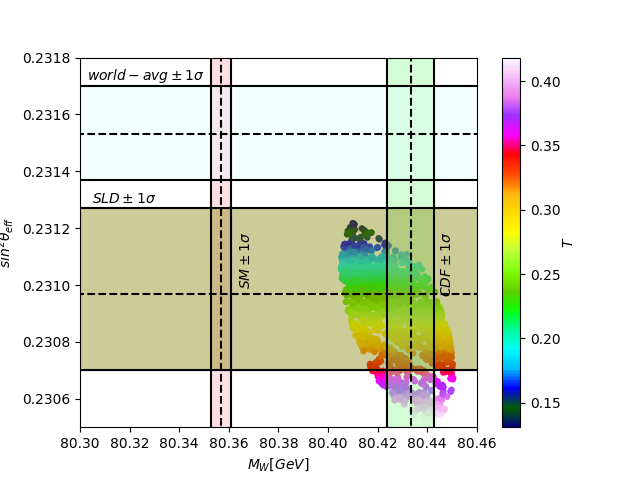}
\\
\hspace*{0.25\textwidth} 
(c) \hfill (d)
\hspace*{0.25\textwidth} 
\caption{Left panels: The $m_{W}$ estimation in the \ott model as a function of $T$, where the color coding represents the size of the parameter $S$. The light pink bands indicate the SM prediction, while the green bands indicate the new CDF measurement for $m_W$, within the 1$\sigma$ uncertainty. Right panels: The $\sin^2\theta_{\rm{eff}}$ prediction in the \ott model as a function of $m_{W}$, with the color coding indicating the size of $T$. The light blue band shows the world average value for $\sin ^2\theta_{\rm{eff}}$ with the associated 1$\sigma$ uncertainty while the brown region illustrates the result from SLD collaboration at the $1\sigma$ level. The upper (lower) panels are for the PDG (CDF) fit result.}
\label{figs_MW_TS}		
\end{figure}
\begin{itemize} 
\item Electro-weak precision observable (EWPO) through the oblique parameters $S$ and $T$ (fixing $U=0$) using both PDG \cite{ParticleDataGroup:2020ssz} and CDF \cite{CDF:2022hxs} fit results. We applied, indeed, the test $\chi^2_{ST}$ before to and following the new $m^{\rm{CDF}}_W$ measurement, indicated by  "PDG" and "CDF", respectively,
\begin{eqnarray}
&&\rm{PDG}: S = 0.05 \pm 0.08,\  T = 0.09 \pm 0.07,\ \rho_{ST} = 0.92\\
&&\rm{CDF}: S = 0.15 \pm 0.08,\  T = 0.27 \pm 0.06,\ \rho_{ST} = 0.93,
\end{eqnarray}
where $\rho_{ST}$ represents the correlation between $S$ and $T$.
\end{itemize} 

\begin{figure}[!hb]
\centering
\hspace*{0.25cm}
\includegraphics[width=0.45\textwidth]{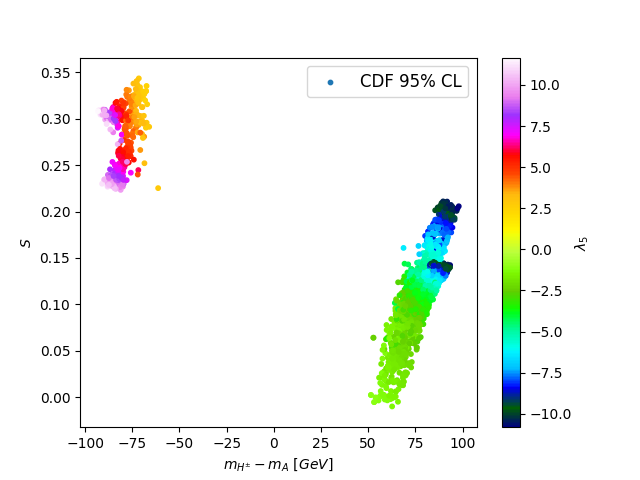}%
\includegraphics[width=0.45\linewidth]{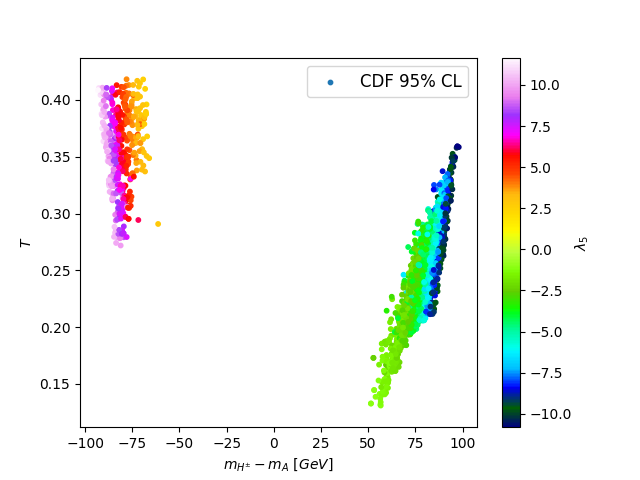}
\\
\hspace*{0.25\textwidth} 
(a) \hfill (b)
\hspace*{0.25\textwidth} 
\\
\hspace*{0.25cm}
\includegraphics[width=0.45\linewidth]{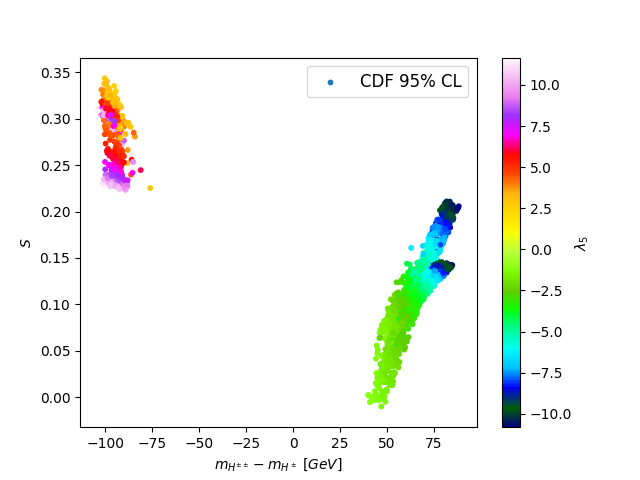}%
\includegraphics[width=0.45\linewidth]{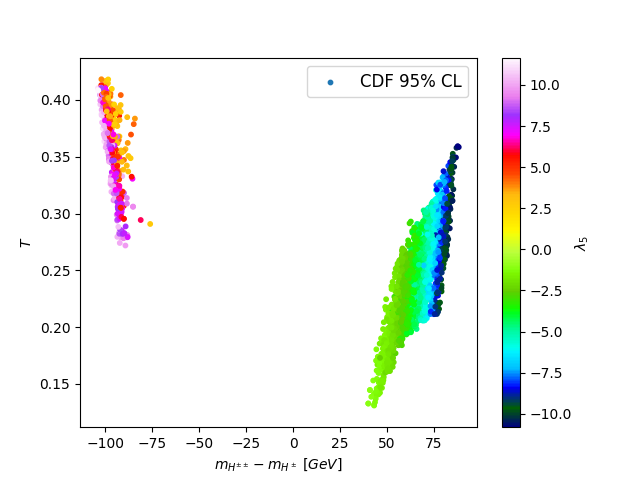}
\\
\hspace*{0.25\textwidth} 
(c) \hfill (d)
\hspace*{0.25\textwidth} 
\caption{Dependence of the $S$ (left) and $T$ (right) oblique parameters on the mass splittings $m_{H^\pm}-m_{A}$ and $m_{H^{\pm\pm}}-m_{H^\pm}$ with the colour code indicating the $\lambda_5$ coupling. The colored regions correspond to the CDF measurement at $95\%$, in full consistency with theoretical and experimental constraints.}
\label{ST_spi}
\end{figure}

The initial summary results are exhibited in Figure.\ref{figs_MW_TS} that qualitatively shows the \ott model loop contributions required by $m_W$ measurement in view of the above PDG and CDF assessment for the oblique parameters. Firstly, by considering the PDG values, we illustrate in Figure.\ref{figs_MW_TS}-(a) the \ott prediction for $m_W$ as a function of $T$ mapped over the $S$ parameter, where the two light pink and green bands show the SM prediction and the new CDF measurement for $m_{W}$ within the 1$\sigma$ uncertainty.  At first sight, it seems clear that the $m_W$ value predicted by the 123-model (while passing all theoretical and experimental constraints discussed briefly above) is in line with the SM prediction, requiring $T \in$ [0.02, 0.10] and $S\in$[0, 0.15], while are so far from the new CDF region at the $1 \sigma$ level. 
However, the CDF bands, if $\chi^2_{STU}({\rm PDG})$ is switched to $\chi^2_{STU}({\rm CDF})$ in the global $\chi^2$, can be construed within the \ott model, thereby enabling the Peskin parameters $T$ and $S$ to slightly lie in the ranges $0.15-0.42$ and $0-0.35$ respectively as can be seen in Figure.\ref{figs_MW_TS}-(c).

On the other side, the right panel in Figure.\ref{figs_MW_TS} exhibits the model prediction for $m_W$ with respect to $\sin^2\theta_{\rm{eff}}$. In such illustration, the light brown and cyan regions show respectively the SLD and world average measurement of $\sin^2\theta_{\rm{eff}}$ at the $1\sigma$ uncertainty. As depicted in Figure.\ref{figs_MW_TS}-(d), after considering the CDF $S$, $T$ result, the parameter points are in good compliance with only SLD measurement for $\sin^2\theta_{\rm{eff}}$ within the $1\text{--}1.5\ \sigma$ level which is not the case when using the PDG values, where the measured value matches well both experiment predictions. 

Thereafter, we will examine how broadly the new CDF $m_W$, and the corresponding $S$ and $T$ parameters would largely affect the mass splitting between the \ott scalars, which is further illustrated in Figure.\ref{ST_spi}, where CDF $S$ and $T$ measurements were considered for the sake of investigation. Clearly, then, it is evident that a consistent prediction of the W boson mass in the range measured by the CDF Collaboration requires sizable mass splittings $m_{H^{\pm}}-m_{A}$ and $m_{H^{\pm\pm}}-m_{H^{\pm}}$ suggesting that such a $m_W$ anomaly can be explained when there is a non-zero mass splitting among the $H^{\pm\pm}$, $H^{\pm}$, $A$. 
Though, in order to peer into these splittings, we've rewrite Eq.(\ref{diffchdmass}) into simple form as,
\begin{equation}
\Delta m^2 \approx m_{H^{\pm\pm}}^2-m_{H^{\pm}}^2 \approx m_{H^{\pm}}^2-m_{A}^2 \approx \frac{1}{2}(m_{H^{\pm\pm}}^2-m_{A}^2) \approx -\frac{\lambda_5}{4}v_\phi^2.
\label{diffchdmass2}
\end{equation}
from which it is fairly evident that such splittings rely mainly on the $\lambda_5$ sign. Hence, for $\lambda_5<0$ the mass difference either between $H^\pm$ and $A$ or $H^{\pm\pm}$ and $H^\pm$ shows the same positive sign, and predict the following hierarchy : $m_{H^{\pm}}-m_{A} > m_{H^{\pm\pm}}-m_{H^{\pm}}$ with a mass splitting ranging from 40 up to 90 GeV among the triplet components, while for $\lambda_5>0$, both splittings are negative, lying between -100 and -66 GeV. Also, it's worth mentioning that at $2\sigma$ of CDF measurement, the values of $ -0.82 \lesssim \lambda_5 \lesssim 2.41$ are excluded, which therefore explains the not allowed mass splittings between roughly -75 GeV and 50 GeV.

\begin{figure}[t]
\centering
\hspace*{0.25cm}
\includegraphics[width=0.45\textwidth]{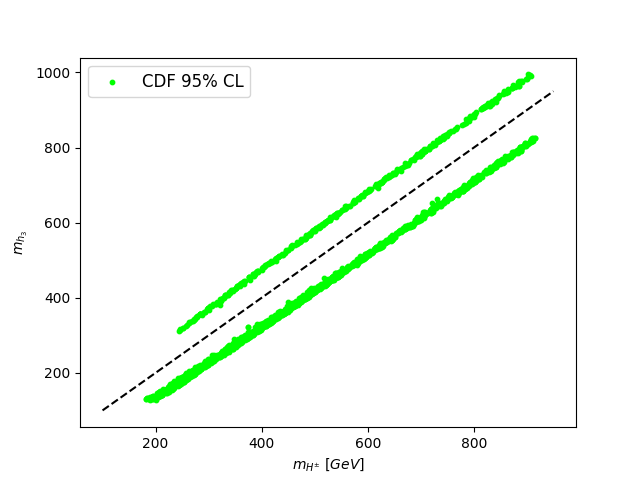}%
\includegraphics[width=0.45\linewidth]{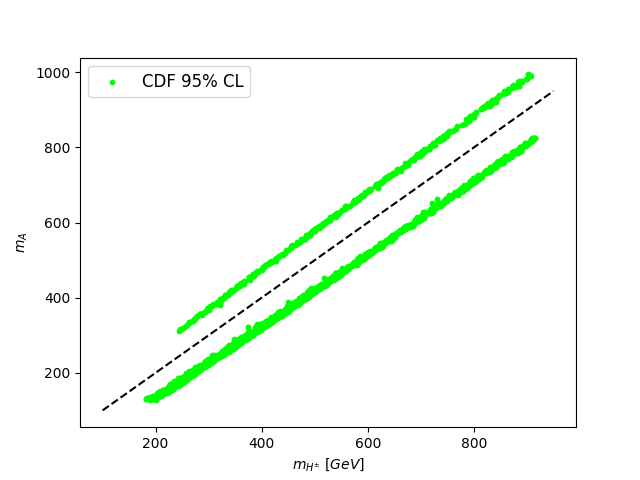}
\\
\hspace*{0.25\textwidth} 
(a) \hfill (b)
\hspace*{0.25\textwidth} 
\\
\hspace*{0.25cm}
\includegraphics[width=0.45\linewidth]{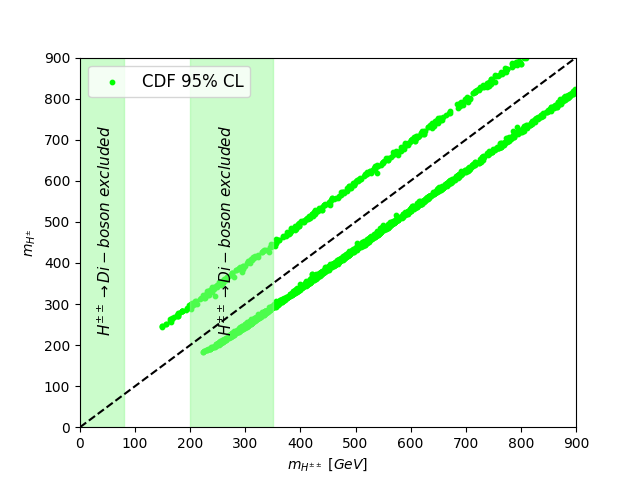}%
\includegraphics[width=0.45\linewidth]{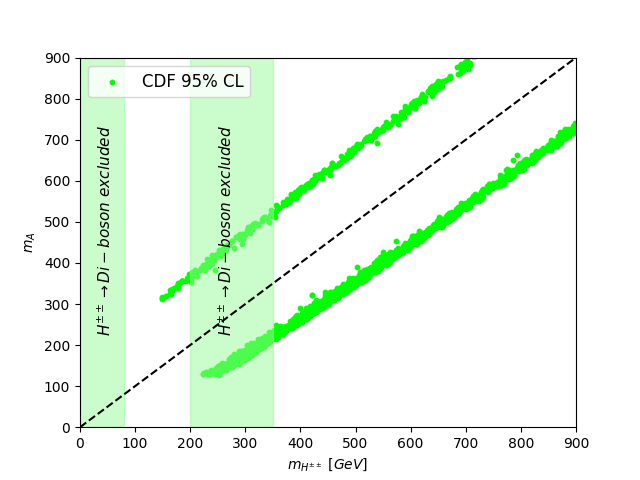}
\\
\hspace*{0.25\textwidth} 
(c) \hfill (d)
\hspace*{0.25\textwidth} 
\caption{Dependence of the Higgs bosons masses on each other on the light of the recent CDF (green points) measurement. The black line indicates the region of the full degenerate masses. The green bands indicate the region excluded by the $H^{\pm\pm} \to W^\pm W^\pm$.}
\label{figs_masses_dep}
\end{figure}

\noindent
To expand a little further on this point, we exhibit in Figure.\ref{figs_masses_dep} the correlation between the Higgs boson masses within two standard deviations of the CDF measurement. Off hand, one could notice how the diagonal for equal masses splits the allowed region in two, which can be expounded by the favored non-zero mass splitting as show in Figure.\ref{ST_spi}. And more importantly, in the \ott model, as it can be seen from Figure.\ref{figs_masses_dep}-(c) and (d), the embedded Higgs bosons are allowed to be as wide below TeV scale, whether it is for $H^\pm$, $H^{\pm\pm}$, $A$, and also $h_{2,3}$, which distinguishes the \ott-model compared to others extensions. By way of example, these large contributions are severely constrained  in HTM. Thus despite the latter is crucial, in imposing also non-zero mass splitting among new particles to express the anomaly in the W boson mass, so long as $v_\Delta$ of order GeV, the neutral heavy Higgs $H$ and single charged Higgs $H^{+}$ masses are required to be low (450 GeV), even 350 GeV for the doubly charged scalar $H^{++}$ \cite{Kanemura:2022ahw}.

Additionally, it is crucial to consider the requirements arising from collider searches. Thus, to cater to the newly measured $m_W$ within the \ott-model, we consider that $\vt \preceq 1$ GeV, such that $H^{\pm\pm}$ could either contribute to Lepton Favor Violation \cite{Gluza:2020qrt}, dominantly decay into same-sign lepton pairs, or two same-sign W bosons. We summarize below the experimental relevant assessments : 
\begin{itemize}
\item [$\circ$] For the doubly charged Higgs boson, $H^{\pm\pm}$, the latest case constitutes the main decay channel \cite{Kang:2014lwn}
\begin{equation}
\Gamma(H^{\pm\pm} \to W^{\pm}W^{\pm}) = \frac{g^2\vt^2m^3_{H^{\pm\pm}}}{16\pi{m_W^2}}\Bigg(\frac{1}{4}-\frac{m_W^2}{m^2_{H^{\pm\pm}}}+\frac{3m_W^4}{m^4_{H^{\pm\pm}}}\Bigg)\,\sqrt{1-4\frac{m_W^2}{m^2_{H^{\pm\pm}}}}
\end{equation}
and a lower bound has been revised \cite{ATLAS:2014kca,Kanemura:2014ipa} to be $m_{H^{\pm\pm}}\ge 84$ GeV, while studying $H^{\pm\pm}$ in the $4W$ final state rules out the $m_{H^{\pm\pm}}$ (in GeV) from 200 up to 350 \cite{ATLAS:2021jol}, as signaled in  Figure.\ref{ST_spi}-(c) and (d). However, in the remaining mass window between 84 and 200 (in GeV), the $H^{\pm\pm}$ is the lightest compared to all the \ott-model Higgs bosons; i.e., $m_{A/h_{2,3}} \ge m_{H^{\pm}} \ge m_{H^{\pm\pm}}$, so that the $l^{\pm}l^{\pm}$ di-lepton decay could have gotten close to the same sign di-boson one, $W^{\pm{\ast}}W^{\pm{\ast}}$. 
\item [$\circ$] For the remaining Higgs bosons, it is also very important to note that situation of degenerate $H^\pm$ with either $h_2$ or $h_3$ is slightly unfavorite by the CDF measurement. More so, only the case were $m_{H^\pm}-m_{h2}>0$ and $m_{H^\pm}-m_{h3}<0$ is allowed. It is noteworthy that CDF measurement has also imposed stringent restrictions on CP odd Higgs boson $m_A$, thus avoiding a possible degeneration of mass with the simply charged Higgs boson ${H^\pm}$, by preferring a positive splitting : $m_{H^\pm}-m_{A} > 0$. 
\end{itemize} 

\section{Conclusion} \label{conclusion}
\hspace*{6mm}
The CDF II experiment has reported a significant anomaly for the $W$ boson mass. And, with remarkable precision, it reveals a slightly higher value compared to the SM value. This intriguing deviation continues to be actively investigated, as it could potentially signify the presence of new physics BSM. It is also plausible that such discrepancy stems from systematic uncertainties or other contributing factors rather than new physics. Further in-depth studies are necessary to fully comprehend the origin of this deviation and its implications in the realm of particle physics. 

In this paper, we have discussed the consistency of the aforementioned anomaly within the 123-model, while considering the theoretical and experimental constraints. Accordingly, as the provided Higgs spectrum is phenomenologically diverse, the impacts of the fields involved VEVs $\vs$ and $\vt$ at the tree level could affect on $S$ and $T$ oblique parameters, and thus can dramatically change the $W$-boson mass from what the SM predicts. We found that, the new $m_W$ measured at CDF-II experiments favorite non-zero mass splitting among the $h_2,\,h_3,\,A, H^\pm$ and $H^{\pm\pm}$. 

All the above, it is noteworthy that the precise measurements of the $W$ boson mass made by CDF, and many others experiments, continue to play a crucial part in putting the SM to the test and in the quest for novel physics outside of the SM. 

\bibliographystyle{JHEP}
\bibliography{bibliography}
\end{document}